# Beyond Boltzmann: The Potential Energy Distribution of Objects in the Atmosphere

Henry Hess

*Department of Biomedical Engineering, Columbia University, USA*

*hhess@columbia.edu*

ABSTRACT. Estimates of the number and potential energy of molecules, aerosols, cloud droplets, insects, birds, planes and satellites in the atmosphere yield a distribution which is for potential energies below $10^2$ $k_BT$ described by the Boltzmann distribution, but for the range from $10^2$ $k_BT$ to $10^{33}$ $k_BT$ by a power law with an exponent of approximately -1. An explanation for this surprising behavior is not found.

Feynman begins his discussion of the principles of statistical thermodynamics with the exponential atmosphere resulting from the Boltzmann law.[1] He then emphasizes that due to a non-uniform temperature and wind the actual atmosphere is neither exponential nor does the ratio of light and heavy molecules at different heights correspond to the prediction from Boltzmann's law.

As one contemplates this in the relative discomfort of economy class of a jet plane in flight, one cannot help but being struck by the fact that at any given moment on the order of 10,000 planes with a mass on the order of 100,000 kg are present at a height of 10,000 m in the atmosphere. These planes have a potential energy of $10^{30}$ $k_BT$, and we would expect from rote application of Boltzmann's formula zero planes to ever leave the ground. Clearly, the atmosphere does not care for thermodynamic equilibrium as it supports this state of affairs. Thus, the question arises what the distribution of the potential energies of objects in the atmosphere actually is.

We can make rough estimates for the abundance and potential energies of different classes of objects:
- Molecules ($10^{44}$ according to the COSPAR International Reference Atmosphere) roughly distributed in potential energy according to the Boltzmann distribution;
- Aerosol particles ($3*10^{27}$ particles with a diameter larger than 10 nm with a potential energy of 2,500 $k_BT$ and $1*10^{27}$ particles with a diameter larger than 100 nm and a potential energy of $10^6$ $k_BT$ using particle densities and height profiles from Watson-Parris et al.[2] and a density of 2 g/cm$^3$);
- Cloud droplets ($4*10^{25}$ particles with an energy of $3*10^{10}$ $k_BT$ using 50 cm$^{-3}$ as average particle density, 10 km as average height, 1 g/cm$^3$ as density, and 2 μm as diameter);
- Insects in flight ($10^{15}$ insects with a potential energy of $3*10^{15}$ $k_BT$ using an estimate of $10^{17}$ flies according to McAlister[3] spending 1% of their time in flight out of a total of $10^{19}$ insects according to *https://www.si.edu/spotlight/buginfo/bugnos*, with a mass of 10 mg and a height of flight of 1 m);
- Birds in flight ($3*10^9$ birds with a potential energy of $3*10^{21}$ $k_BT$ using an estimate of 300 billion birds according to Gaston and Blackburn[4] spending an estimated 1% of their time in flight at a height of 10 m with an average weight of 100 g);
- Small planes and helicopters in flight (1000 with a potential energy of $3*10^{27}$ $k_BT$ using an estimated number of $10^5$ planes with a weight of 1,000 kg spending 1% of their time at an altitude of 1,000 m, see e.g. *https://en.wikipedia.org/wiki/Cessna_172*);
- Jet planes in flight (5,000 planes with a potential energy of $3*10^{30}$ $k_BT$ using an estimated number of 25,000 operating planes with a weight of 80,000 kg spending 20% of their time at an altitude of 10,000 m, *https://en.wikipedia.org/wiki/List_of_jet_airliners*);
- Satellites in low-Earth orbit (3,000 satellites with a potential energy of $5*10^{30}$ $k_BT$ using a weight of 1,000 kg and a height of 2,000 km, see *https://en.wikipedia.org/wiki/Satellite*);

- Satellites in medium-Earth orbit (300 satellites with a potential energy of 10^31 $k_BT$ using a weight of 1,000 kg and a height of 20,000 km, see *https://en.wikipedia.org/wiki/Satellite*);
- Satellites in geostationary orbit (500 satellites with a potential energy of 10^31 $k_BT$ using a weight of 1,000 kg and a height of 40,000 km, see *en.wikipedia.org/wiki/Geosynchronous_satellite*);
- The International Space Station (with a potential energy of 2*10^32 $k_BT$ using a weight of 400 tons and a height of 400 km, see *en.wikipedia.org/wiki/International_Space_Station*).

These estimates are summarized in Figure 1 where the cumulative number of objects of a given minimum energy is plotted on a log-log plot to capture the long tail of the distribution. While the potential energy distribution of the molecules follows the Boltzmann distribution on this log-log plot, larger objects form a long tail.

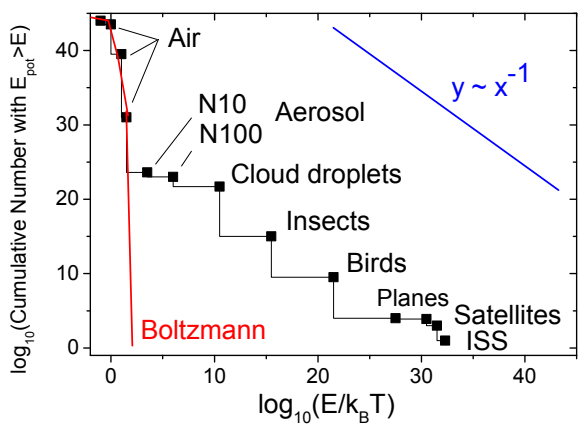

**Figure 1:** The cumulative number of objects in the atmosphere with an energy exceeding the energy on the x-axis.

While the estimates obviously can easily have errors of 1-2 orders of magnitude in any direction, and the actual distribution is obviously much smoother due to the variability of the mass and height of the objects in each class, the "Gestalt" of the tail of the distribution is surprisingly close to a power law with an exponent of -1. It is not immediately obvious why this should be the case.

Two potentially significant contributors to the tail of the distribution which are not accounted for here are large hailstones, whose size-dependent frequency the author found impossible to estimate on a global basis, and the frequency of "temporarily captured objects" from outer space, which potentially exceeds the contribution of planes and satellites according to estimates of their abundance by Granvik et al.[5].

It is worth asking if the type of movement is the key difference between the small molecules obeying a Boltzmann distribution and the larger objects forming the long tail. It may be also relevant that collisions between molecules generally do not lead to permanent alterations of the molecules, whereas collisions between larger objects are frequently followed by coalescence (raindrops), destruction (planes), or destructive coalescence (plane and bird), but rarely by fragmentation into smaller objects of the same type. Avoidance of objects of similar size may be necessary, but it is not obvious how this translates into a distribution of potential energy or a potentially underlying distribution of density.

In search of a mechanistic explanation, the mind is also drawn to similar power law distribution in biology[6] or technology[7], or the rich body of work on metabolic ecology[8,9]. However, the connection between cloud droplets, animals, and machines appears on first glance elusive.

In summary, actually looking at what is in the atmosphere raises new and puzzling questions.